\documentclass[aps,showpacs,nofootinbib,twocolumn,preprintnumbers,nobalancelastpage,superscriptaddress]{revtex4}

\usepackage[english]{babel}
\usepackage{amsmath,amssymb}
\usepackage[dvips]{graphicx}
\usepackage[sort&compress]{natbib}
\usepackage{amsfonts}
\usepackage{bbm}
\usepackage{color}

\DeclareMathAlphabet{\mathsc}{OT1}{cmr}{m}{sc}

\allowdisplaybreaks

\def\10{$SO(10)$}
\def\21{SU(2) $\otimes$ U(1) }

\def\422{$SU(4) \otimes SU(2) \otimes SU(2)$}
\def\321{SU(3) $\otimes$ SU(2) $\otimes$ U(1)}

\def\lsim{\raise0.3ex\hbox{$\;<$\kern-0.75em\raise-1.1ex\hbox{$\sim\;$}}}
\def\gsim{\raise0.3ex\hbox{$\;>$\kern-0.75em\raise-1.1ex\hbox{$\sim\;$}}}

\newcommand{\flux}[2][]{\ensuremath{\ifthenelse{\equal{#1}{}}{}{^{#1}\!}\mathit{#2}}}

\newcommand{\AddrAHEP}{%
  AHEP Group, Instituto de F\'{\i}sica Corpuscular --
  C.S.I.C./Universitat de Val{\`e}ncia \\
  Edificio Institutos de Paterna, Apt 22085, E--46071 Valencia, Spain}

\begin{document}


\title{Finding the Higgs Boson through Supersymmetry}

\author{F.\ de Campos}
\email{camposc@feg.unesp.br}
\affiliation{Departamento de F\'{\i}sica e Qu\'{\i}mica,
             Universidade Estadual Paulista, Guaratinguet\'a -- SP,  Brazil }

\author{O.\ J.\ P.\ \'Eboli}
\email{eboli@fma.if.usp.br}
\affiliation{Instituto de F\'{\i}sica,
             Universidade de S\~ao Paulo, S\~ao Paulo -- SP, Brazil.}

\author{M.\ B.\ Magro}
\email{magro@fma.if.usp.br}
\affiliation{Instituto de F\'{\i}sica,
             Universidade de S\~ao Paulo, S\~ao Paulo -- SP, Brazil.}
\affiliation{Centro Universit\'ario Funda\c{c}\~ao Santo Andr\'e,
             Santo Andr\'e -- SP, Brazil.}

\author{D.\ Restrepo}
\email{restrepo@uv.es}
\affiliation{\AddrAHEP}
\affiliation{Instituto de F\'{\i}sica, Universidad de Antioquia - Colombia}

\author{J. W. F. Valle}
\email{valle@ific.uv.es}
\affiliation{\AddrAHEP}

\begin{abstract}
  The study of displaced vertices containing two b--jets may provide a double
  discovery at the Large Hadron Collider (LHC): we show how it may not only
  reveal evidence for supersymmetry, but also provide a way to uncover the
  Higgs boson necessary in the formulation of the electroweak theory in a
  large region of the parameter space. We quantify this explicitly using the
  simplest minimal supergravity model with bilinear breaking of R--parity,
  which accounts for the observed pattern of neutrino masses and mixings seen
  in neutrino oscillation experiments.  
  \end{abstract}


\maketitle


\section{Introduction}

By opening the exploration of the new territory of physics at the
Terascale, the CERN Large Hadron Collider (LHC) is likely to shed
light upon the main open puzzle in particle physics, namely the origin
of mass and the nature of electroweak symmetry breaking.
Supersymmetry (SUSY) provides an elegant way of justifying the
electroweak symmetry breaking mechanism in terms of an elementary
Higgs particle, alleviating the so called hierarchy
problem~\cite{Dimopoulos:1981yj}.
The Higgs boson and the existence of supersymmetry therefore stand out
as the main missing pieces in our understanding of fundamental forces,
and a lot of effort has been put into their direct observation.
Indeed the search for the Higgs boson and for supersymmetry constitute
the main topic in the agenda of the LHC. \medskip

In contrast, so far the only established evidence for physics beyond
the standard model (SM) has been the discovery of neutrino masses and
oscillations~\cite{Maltoni:2004ei}, which has culminated decades of
painstaking efforts. \medskip

Here we stress that these two issues may be closely related. Indeed,
low-energy supersymmetry with broken R--parity~\cite{Hall:1983id}
provides a plausible mechanism for the origin for neutrino masses and
mixings. Indeed, as the bilinear model best
illustrates~\cite{Hirsch:2004he}, in contrast to the simplest seesaw
schemes~\cite{Nunokawa:2007qh}, these may be tested at particle
accelerators like the LHC~\footnote{ { Such model has no conventional
    neutralino dark matter, though other possible dark matter
    candidates may be envisaged such as the
    axion~\cite{Peccei:1977hh}, the majoron~\cite{Berezinsky:1993fm},
    the axino or the gravitino~\cite{Hirsch:2005ag}.} }.\medskip

Here we consider the simplest ansatz to introduce R--parity breaking
in supersymmetry, characterized by an additional bilinear violating
(BRpV) term in the superpotential~\cite{Hirsch:2000ef}. It provides
the simplest effective description of a more complete picture
containing additional neutral heavy lepton~\cite{Dittmar:1989yg}
superfields whose scalars drive the spontaneous breaking of
R--parity~\cite{Masiero:1990uj}. \medskip

Our focus here is on the specific case of a minimal gravity mediated
supersymmetry breaking model with bilinear R parity violation:
BRpV--mSUGRA model for short. In this model, the lightest
supersymmetric particle (LSP) is no longer stable. Current neutrino
oscillation data indicate that the strength of the BRpV term is
small~\cite{Hirsch:2000ef}, hence the LSP decay length is expected to
be long enough to provide a displaced vertex at the
LHC~\cite{deCampos:2005ri,deCampos:2007bn}.  { For a low Higgs mass
  the dominant decay is into { $b\bar{b}$}, however at the LHC the
  overwhelming QCD background makes this signal irrelevant when the
  Higgs is produced in the standard way.
  In supersymmetry the Higgs can be produced after the decay chains of
  the next-to-lightest supersymmetric particle. In the R--parity
  conserving case for specific spectrum and supersymmetric production,
  the additional jets and the missing energy can allow the discovery
  of the Higgs in the $b$ channel \cite{19}. The same features also
  hold in our case, but in addition now the Higgs can be produced from
  the lightest neutralino, leading to events with a displaced vertex {
    with two large invariant mass b--jets}. The signal of a neutralino
  into a Higgs and a neutrino is therefore free of SM backgrounds if
  the neutralino decays inside the pixel detector and well outside the
  interaction point. Here we show explicitly that this is the
  case~\footnote{In fact the LHCb collaboration is considering the
    possibility to search for b's originating outside the interaction
    point \cite{20}.}.} \medskip

In this work we analyze the potential of the LHC to survey the
existence of the Higgs boson using a novel signal: a b--jet pair
coming from displaced vertices generated by the lightest neutralino
decays within the BRpV--mSUGRA model.  We demonstrate that the LHC
reach is capable to uncover a supersymmetric Higgs in a fair region of
the $M_{1/2}\otimes M_0$ parameter plane. \medskip


\section{{ Model description}}

The  BRpV model is described by the superpotential
\begin{equation}
W_{\text{BRpV}} = W_{\text{MSSM}}  + \varepsilon_{ab}
\epsilon_i \widehat L_i^a\widehat H_u^b \; ,
\end{equation}
in which the standard minimal supersymmetric model (MSSM) is
supplemented by three extra bilinear terms characterized by three new
parameters ($\epsilon_i$), one for each fermion generation. In
addition to these we must also include new soft supersymmetry breaking
terms ($B_i$) in whose presence the bilinears become physical
parameters that can not be rotated away~\cite{Diaz:1997xc}.
\begin{equation}
V_{\text{soft}} = V_{\text{MSSM}} - \varepsilon_{ab} 
B_i\epsilon_i\widetilde L_i^aH_u^b
\end{equation}
The new terms in the BRpV Lagrangian ($\epsilon_i$, $B_i$) lead to the
explicit violation of lepton number as well as R--parity. Furthermore,
the sneutrino fields acquire a vacuum expectation value when we
minimize the scalar potential. \medskip

In BRpV models the terms presenting explicit lepton number violation,
as well as, the sneutrino vacuum expectation values generate mixing
among neutrinos and neutralinos giving rise to one tree--level
neutrino mass. The other two neutrino masses are generated through
loop diagrams~\cite{Hirsch:2000ef}. One can show that, indeed, the
resulting neutrino masses and mixings provide a good description of
all current neutrino oscillation data~\cite{Maltoni:2004ei}. \medskip

For the sake of definiteness, we assume mSUGRA as the model of
supersymmetry breaking, implying universality of the soft breaking
terms at unification. In this case, our model depends upon eleven free
parameters, namely
\begin{equation}
M_0\,,\, M_{1/2}\,,\, \tan\beta\,,\, {\mathrm{sign}}(\mu)\,,\,
A_0 \,,\,
\epsilon_i \: {\mathrm{, and}}\,\, \Lambda_i\,,
\end{equation}
where $M_{1/2}$ and $M_0$ are the common gaugino mass and scalar soft
SUSY breaking masses at the unification scale, $A_0$ is the common
trilinear term, and $\tan\beta$ is the ratio between the Higgs field
vev's. For convenience, we trade the soft parameters $B_i$ by
$\Lambda_i=\epsilon_iv_d+\mu v_i$, where $v_i$ is the vacuum
expectation value of the sneutrino fields, since the $\Lambda_i$'s are
more directly related to the neutrino masses; for further details see
~\cite{Hirsch:2000ef}. \medskip

The bilinear R--parity violating interaction gives rise to mixings
between SM and SUSY particles that lead to decay of the LSP into SM
particles. In a large fraction of the parameter space the lightest
neutralino is the LSP and it can decay into leptonic final states $\nu
\ell^+ \ell^{\prime -}$, where $\ell = e$, $\mu$ or $\tau$, as well as
into semi-leptonic final states $ \ell q^\prime \bar{q}$ or $ \nu q
\bar{q}$. For sufficiently heavy neutralinos these decays are
dominated by two--body channels like $\nu Z$, $\ell^\pm W^\mp$ and
$\nu h$ with $h$ being the lightest CP even Higgs boson; for further
details see~\cite{Porod:2000hv,deCampos:2007bn,deCampos:2008re}. In
the region where the stau is the LSP the detached vertex signal
disappears completely since the stau possesses a very small decay
length. \medskip

In contrast, a salient feature of our BRpV model is that neutralino
LSPs exhibit a rather large decay length, ranging from a few
millimeters to tenths of millimeters for $M_{1/2}$ varying from 200
GeV to 1 TeV. Such large decay lengths lead to the production of
detached vertices at the LHC which constitute a smoking gun of this
kind of models. \medskip

In this work, we analyze the two--body lightest neutralino decay into
the lightest Higgs boson $h^0$ as a Higgs discovery channel
\begin{equation}
\tilde{\chi}^0_1 \to h \nu\;\;\;.
\end{equation}
If the lightest neutralino lives long enough it will be detached from
the primary interaction point leaving a displaced vertex as signal at
the LHC. Since the Higgs boson $h$ decays mostly into a b--quark pairs
we expect a displaced vertex with two b--jets as a characteristic
signature for Higgs production. \medskip

\begin{figure}
  \begin{center}
  \includegraphics[width=0.98\linewidth]{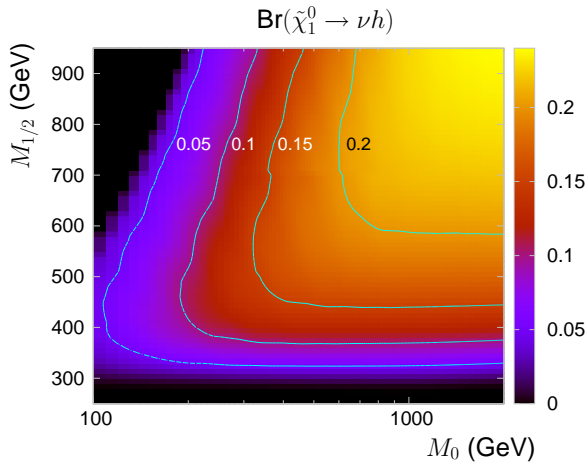}
  \end{center}
  
\vspace{-0.5cm}
\caption{Br$(\tilde\chi_1^0\to h\nu)$ as a function of $M_{1/2}
  \otimes M_0$ for $\tan\beta=10$, $A_0=-100$ GeV and $\mu>0$.}
  \label{chitohnu}
\end{figure}

We present, in Figure \ref{chitohnu}, the lightest neutralino
branching ratio to $h\nu$ as a function of $M_{1/2} \otimes M_0$ for
$\tan\beta=10$, $A_0=-100$ GeV and $\mu>0$~\footnote{ We note
    that in the upper left dark region the stau is the LSP and in what
    follows we will not consider this region.}.
Here we focus on the situation where the lightest neutralino is
heavier then $h$, so the neutralino Higgs decay channel opens for
$M_{1/2} \gtrsim {\cal O} (300)$ GeV for our choice of parameters.
The maximum value of the branching ratio for this channel is about
$22\%$; for an illustration of the full behavior of neutralino decays
see, for example,
Ref.~\cite{Porod:2000hv,deCampos:2007bn,deCampos:2008re}.  This figure
tells us that, for fixed values of $M_{1/2}$, the LSP branching ratio
into Higgs--neutrino pairs initially grows with increasing $M_0$,
stabilizing for $M_0$ in excess of a few hundred GeV.  On the other
hand, the importance of this decay increases with $M_{1/2}$ for
moderate and large values of $M_0$.
%

%

\section{Signal and backgrounds}


In order to simulate the Higgs production we calculate all R--parity
violating branching ratios and SUSY spectra using the package
SPheno~\cite{Porod:2003um}. We used PYTHIA version 6.408~\cite{pythia}
to generate events, using the SPheno output in the SLHA
format~\cite{Skands:2003cj}.  In order to have a rough simulation of
the detector response we smeared the track energies, but not their
directions, with a Gaussian error given by $\Delta E/E = 0.10/\sqrt{E}
+ 0.01$ (E in GeV) for leptonic tracks and $\Delta E/E = 0.5/\sqrt{E}
+ 0.03$ for all hadronic tracks.\medskip


Displaced vertices at the LHC were identified requiring that the
neutralino decays away from the primary vertex point, that is, outside
an ellipsoid centered at the primary vertex
\begin{equation}
      \left ( \frac{x}{{ 5}\delta_{xy}} \right )^2
   +  \left ( \frac{y}{{ 5}\delta_{xy}} \right )^2
   +  \left ( \frac{z}{{ 5}\delta_{z}} \right )^2   = 1 \; ,
\end{equation}
where the $z$-axis is along the beam direction.  To be conservative we
assumed the ellipsoid size to be five times the ATLAS expected
precision in each direction for the semiconductor
tracker~\cite{atlas:2008zzm} which are $\delta_{xy} = 20~\mu$m and
$\delta_z = 500~\mu$m.  To reconstruct the vertices we required that
visible tracks coming from neutralino decays must have an intersection
inside a sphere determined by the tracking detector resolution which
we assumed to be $10~\mu$m~\cite{atlas:2008zzm}.  Furthermore, we
considered only the charged tracks inside the pseudo--rapidity region
of $|\eta| < 2.5$.\medskip

Since the Higgs production in the LSP decay is characterized by the
presence of two b--tagged jets we looked for events with at least one
displaced vertex containing at least one jet tagged as a b--jet. In
our analyses we considered a b--tagging efficiency up to
$50\%$.\medskip

In order to ensure that the detached vertex events are properly
recorded we accepted only events that pass very simple trigger
requirements.We further required the events to present an isolated
electron (muon) with $p_T>20$ (6) GeV, or the presence of a jet with
$p_T > 100$ GeV, or missing transverse energy in excess of 100 GeV.
\medskip

For our analysis we have fixed $\tan\beta = 10$, $A_0 = -100$ GeV and
$\mu > 0$. For this choice of parameters, the Higgs mass lies in the
range $ 110$ GeV $\lsim M_{h} \lsim 120$ GeV when we vary $M_0$ and
$M_{1/2}$.  Since we are only interested in detached jets coming from
Higgs decays, we have further required that the jet--jet invariant
mass is around the Higgs mass value.\medskip

Within the SM framework displaced vertices originate from decays of
long lived particles, like $B$'s and $\tau$'s, and consequently its
visible decay products exhibit a rather small invariant mass.  In
contrast, in our BRpV model, the displaced vertices are associated to
the LSP decay and will have in general a large invariant mass
associated to them.  Therefore, physical SM processes do not lead to
sizeable backgrounds to the detached Higgs searches due to to large
difference in the invariant mass of the visible products.  However,
BRpV LSP decays into $\nu Z$ are a potential source of background for
the Higgs signal.\medskip

\begin{figure}[th]
  \begin{center}
  \includegraphics[width=0.98\linewidth]{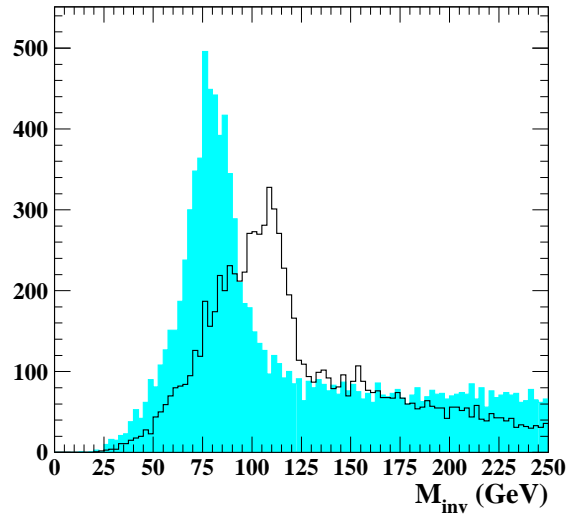}
  \end{center}
  \vspace*{-8mm}
  \caption{Jet pair invariant mass distribution in GeV. The light blue
    (greyish) histogram stands for the background where the lightest
    neutralino decays via W and Z bosons and the other histogram
    stands for the channels where the lightest neutralino decays into
    $b\bar{b}$ pairs.}
\label{wpeak}
\end{figure}

As an illustration we show in Figure \ref{wpeak} the jet--jet
invariant mass distribution of all displaced vertices exhibiting
jets. As we can see, a cut on the invariant mass outside the range
$100$ GeV $< M_{inv} < 125$ GeV eliminates a good fraction of
supersymmetric backgrounds coming, for instance, from the neutralino
decay into W and Z bosons as well as the three--body $b\bar{b}\nu$
channel.  The physical background can be further suppressed by
requiring that at least one of the jets associated to the displaced
vertex is tagged as a b jet.  Moreover, these requirements ensure that
SM backgrounds coming from the decay of long lived particles are also
efficiently eliminated.  There remain instrumental
backgrounds~\cite{Strassler:2008fv} which require a full detector
simulation along the lines we have described above; this simulation is
beyond the scope of the present work. \medskip

\begin{figure}[hb]
  \begin{center}
\includegraphics[width=0.98\linewidth]{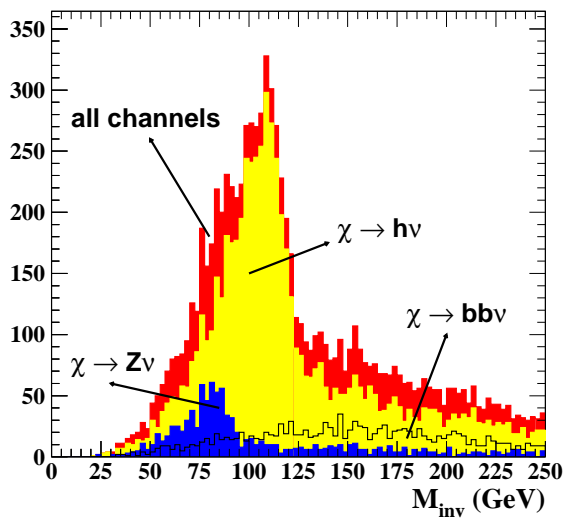}
  \end{center}
  \vspace*{-8mm}
  \caption{Invariant mass distribution in GeV of the neutralino decaying into
  b-jet pairs separated into its several channels.}
\label{minv_cut}
\end{figure}

In Figure~\ref{minv_cut} we show that almost all vertices containing
b--jets come from neutralino decay via Higgs and that our invariant
mass cut will eliminate the $\nu Z$ background, while keeping a large
fraction of the signal events.  We checked that the events passing the
LHC triggers and all the above cuts come from the signal events
$\tilde{\chi}^0_1 \to \nu h$ with the physics background being
negligible.\medskip

In order to estimate the LHC reach for Higgs search coming from
displaced vertex signal in BRpV--mSUGRA models we considered a few
scenarios. In the optimistic analysis we assumed that there is no
event coming from instrumental backgrounds or overlapping events and
took the b--tagging efficiency to be 50\%. In this case we required
that the signal must have more than 5 events since no background is
expected and present our result in the $M_{1/2}\otimes M_0$ plane for
integrated luminosities of 10 and 100 fb$^{-1}$. We also considered
three additional scenarios. In the first one we studied the impact of
a lower b--tagging efficiency (30\%) but we still assumed that the
process is background free. In the second case we assumed that there
are 5 backgrounds events originating from instrumental errors and
overlapping events and required a $5\sigma$ signal for a 50\%
b--tagging efficiency. Finally, in the last scenario we assumed the
same background as in the last case, lowering however, the b--tagging
efficiency down to 30\%. \medskip

\section{Results}

\begin{figure}[b]
  \begin{center}
\includegraphics[width=0.98\linewidth]{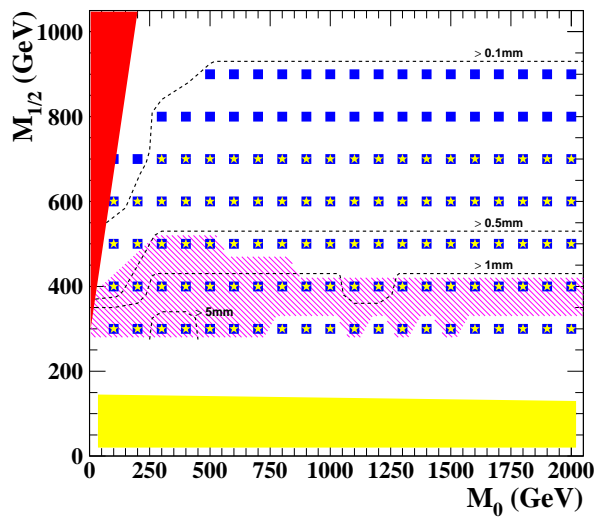}
  \end{center}
  \vspace*{-8mm}
  \caption{LHC reach for Higgs search in displaced vertices for the
    BRpV--mSUGRA model in the plane $M_{1/2}\otimes M_0$ assuming
    $\tan\beta = 10$, $A_0=-100$ GeV, and $\mu > 0$. The yellow stars
    (blue squares) represent the reach for an integrated luminosity of
    10 (100) fb$^{-1}$ while the hatched region corresponds to the
    reach of the LHCb experiment for an integrated luminosity of 10
    fb$^{-1}$.  The (yellow) shaded region in the bottom stands for
    points excluded by direct LEP searches, while the (red)
    upper--left area represents a region where the stau is the
    LSP. Note that the black lines
    delimit different regimes of LSP decay length.}
\label{fig:reach}
\end{figure}

In Figure~\ref{fig:reach} we depict the LHC discovery reach for the
Higgs displaced vertex signal in our most optimistic scenario.
The shaded (yellow) region at the bottom stands for points already
excluded by direct LEP searches while the upper--left corner of the
$M_{1/2}\otimes M_0$ plane, the (red) shaded area, corresponds to the
region where the stau is the
LSP~\cite{deCampos:2007bn}, and hence
is not covered by the present analysis.
The region around $M_{1/2} = 200$ GeV has no signal due to the fact
that the neutralino mass is smaller than the Higgs mass in it,
therefore, being forbidden the two--body LSP decay into
Higgs--neutrino pairs.\medskip

From Fig.~\ref{fig:reach} one can see that the ATLAS and CMS
experiments will be able to look for the signal up to $M_{1/2} \sim
700$ $(900)$ GeV for a LHC integrated luminosity of 10 (100)
fb$^{-1}$.
Notice that the LHC Higgs discovery potential is almost independent of
$M_0$. For a fixed value of $M_{1/2}$ the LSP total production cross
section decreases as $M_0$ increases, however, the LSP branching ratio
into Higgs--neutrino pairs increases with $M_0$, therefore, both
effects tend to cancel and produce the observed behavior.
Moreover, this figure also exhibits the average decay length of the
neutralino, demonstrating that its decay takes place inside the vertex
detector, ensuring a good vertex reconstruction.\medskip

We have also estimated the reach expected at LHCb for our Higgs search
proposal.  The hatched region in Fig.~\ref{fig:reach} indicates the
LHCb reach for 10 fb$^{-1}$. Due to the strong cut on the
pseudo--rapidity required by this experiment the reach for 2 fb$^{-1}$
is severely depleted and only a small region of the parameter space is
covered, {\em i.e.}, $300\; \rm{GeV} \leq M_{1/2} \leq 350$ GeV and
$200\; \rm{GeV} \leq M_{0} \leq 500$ GeV.\medskip

\begin{figure}[t]
  \begin{center}
\includegraphics[width=0.98\linewidth]{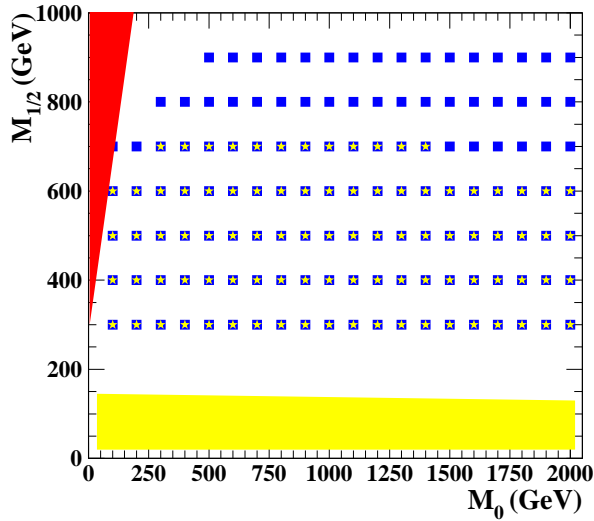}
  \end{center}
  \vspace*{-8mm}
  \caption{Same as Fig.~\ref{fig:reach} using a b-jet reconstruction
    efficiency of 30$\%$ with no background events. }
\label{fig:eff30}
\end{figure}

Tagging b--jets emanating from a detached vertex is certainly a more
intricate procedure, therefore, we also considered a lower b--tagging
efficiency in our analyses.  Figure~\ref{fig:eff30} contains the reach
of LHC for Higgs search using a b--jet reconstruction efficiency of
30$\%$, instead of 50$\%$ used of Fig.~\ref{fig:reach}, however, we
still assumed that the search is background free.  Comparing
Figs.~\ref{fig:reach} and \ref{fig:eff30}, one can see that the LHC
reach in this second case is mildly affected by this change for an
integrated luminosity of 10 fb$^{-1}$, while the changes are minute at
higher integrated luminosities.

\medskip

\begin{figure}[t]
  \begin{center}
\includegraphics[width=0.98\linewidth]{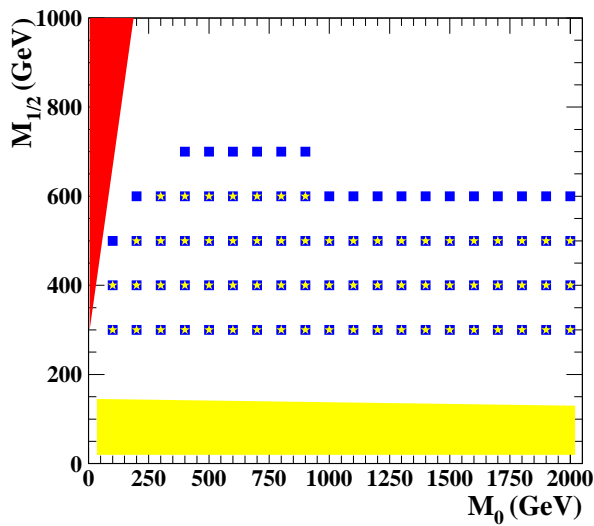}
  \end{center}
  \vspace*{-8mm}
  \caption{Same as Fig.~\ref{fig:reach} using a b--jet reconstruction
    efficiency of 50$\%$ and assuming the existence 5 background
    events for both integrated luminosities.}
\label{fig:5event}
\end{figure}

A study of the instrumental backgrounds and the effect of overlapping
events does require a full detector simulation, which is beyond the
scope of this work.  In order to assess the impact of existence of
non--physical backgrounds we considered that these backgrounds give
rise to 5 background events for both integrated luminosities used in
our studies.  In Figure~\ref{fig:5event} we present the $5\sigma$
LHC Higgs discovery potential assuming a b--jet reconstruction
efficiency of 50$\%$ and 5 background events. We can see from this
figure that the existence of background events does lead to a
substantial reduction of the LHC reach for Higgs in displaced
vertices. \medskip

%

\begin{figure}[thb]
  \begin{center}
\includegraphics[width=0.98\linewidth]{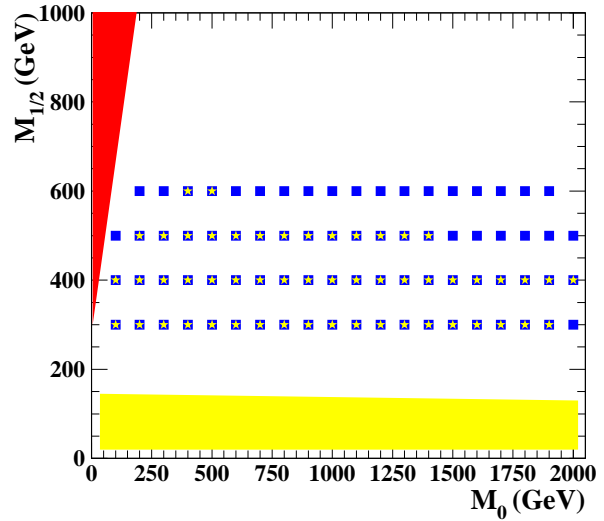}
  \end{center}
  \vspace*{-8mm}
  \caption{Same as Fig.~\ref{fig:reach} using a b-jet reconstruction
    efficiency of 30$\%$ and 5 background events.}
\label{fig:5evt-eff}
\end{figure}

In Fig.~\ref{fig:5evt-eff} we present the reach of LHC for Higgs
search in a very pessimistic scenario that exhibits a lower b--jet
reconstruction efficiency of 30$\%$, as well as, the presence of 5
background events.  In this case we observe a more severe reduction of
the LHC reach that is reduced to $M_{1/2} = 600$ GeV at most.  This
large depletion of the LHC search potential follows from the need of a
large number of signal events to establish the signal given the fast
decrease of the SUSY production cross section with increasing
$M_{1/2}$.  In this sense, the 100 fb$^{-1}$ case is more affected
since the production cross section exhibits a steep decrease for
$M_{1/2} \gtrsim 700$ GeV.\medskip

\section{Conclusions}

In summary we have seen how the search for displaced vertices
containing b--tagged jets at the LHC may not only provide evidence for
supersymmetric particles but also lead to the discovery of the Higgs
boson of the electroweak theory. We have given a quantitative analysis
within the simplest minimal supergravity model with bilinear breaking
of R--parity, which accounts for the observed pattern of neutrino
masses and mixings observed in current neutrino oscillation
experiments.  Similar variant schemes can be envisaged where, for
example, supersymmetry and/or electroweak breaking is realized
differently.\medskip

In an optimistic background free scenario the Higgs search in LSP
decays can be carried out for LSP masses up to 300 (380) GeV for an
integrated luminosity of 10 (100) fb$^{-1}$. We showed that this
result is robust against variations of the assumed b--tagging
efficiencies.  Notwithstanding, the results change drastically if
instrumental backgrounds are present. Assuming the existence of 5
background events reduces the LHC reach to LSP masses of 210 (250) GeV
at the low (high) luminosity run.\medskip


\section*{Acknowledgments}

We thank A. Bartl for careful reading the manuscript. This work was
supported by MEC grant FPA2005-01269, by EC Contracts RTN network
MRTN-CT-2004-503369, by Conselho Nacional de Desenvolvimento
Cient\'{\i}fico e Tecnol\'ogico (CNPq) and by Funda\c{c}\~ao de Amparo
\`a Pesquisa do Estado de S\~ao Paulo (FAPESP) and by Colciencias in
Colombia under contract 1115-333-18740.


\bibliographystyle{h-physrev4}

\end{document}